\documentclass[11pt,a4paper]{article}
\usepackage[margin=1.0 in]{geometry} 

\newcommand{\appref}[1]{\hyperref[#1]{Appendix \ref*{#1}}}
\usepackage[]{caption} 
\usepackage{breakurl}
\usepackage{amssymb}
\usepackage[utf8]{inputenc}
\usepackage[T1]{fontenc} 
\usepackage{graphicx}
\usepackage{mathtools}
\graphicspath{ {images/} }
\usepackage{amsmath}
\usepackage{amsthm}
\usepackage{enumerate}
\usepackage{float,booktabs}
\usepackage[T1]{fontenc}
\usepackage{pdflscape}
 
\usepackage{tabularx}
\usepackage{comment}
\usepackage{titlesec}
\usepackage{dsfont}
\usepackage{nicefrac}
\titleformat{\paragraph}[runin]{\normalfont\normalsize}{\theparagraph}{1em}{}
\titlespacing*{\paragraph}{0pt}{3.25ex plus 1ex minus .2ex}{\the\fontdimen2\font}

\numberwithin{equation}{section} 

\usepackage{amsfonts}
\usepackage{authblk}
\usepackage{subcaption}
\usepackage{geometry}
\usepackage{hyperref}
\usepackage{rotating}
\usepackage[bottom]{footmisc}
\usepackage{lscape}
\usepackage{siunitx}
\usepackage{hhline}
\usepackage{caption}
\usepackage{array} 
\usepackage{esint}
\usepackage[authoryear,longnamesfirst,round]{natbib} 
\usepackage{eurosym}
\usepackage{bm}
\usepackage{soul}
\usepackage{xcolor}

\newtheorem{theorem}{Theorem}

\newtheorem{assumption}{Assumption}
\newtheorem{corollary}{Corollary}

\newtheorem{lemma}{Lemma}

\newtheorem{proposition}{Proposition}

\newtheorem{assumptionalt}{Assumption}[assumption]

\newenvironment{assumptionp}[1]{
  \renewcommand\theassumptionalt{#1}
  \assumptionalt
}{\endassumptionalt}

\usepackage{amsthm}

\newtheorem{subassumption}{Assumption\normalfont}[assumption]

\newenvironment{assumption*}
 {\ifnum\value{subassumption}=0 \stepcounter{assumption}\fi\subassumption}
 {\endsubassumption}
\newenvironment{assumption+}[1]
 {\renewcommand{\thesubassumption}{#1}\subassumption}
 {\endsubassumption}

\newcommand{\expo}{\text{exp}}

\newcommand{\pltilde}{\tilde{P}^L_{it}}
\newcommand{\dpar}[2]{\frac{\partial #1}{\partial #2}}

\newcommand*{\prob}{\mathsf{P}}

\hypersetup{
  colorlinks,
  allcolors=[rgb]{0.0, 0.0, 0.5}
  }

\allowdisplaybreaks 

\title{\LARGE{On the Non-Identification of Revenue Production Functions}} 

\author{David Van Dijcke\thanks{\texttt{dvdijcke@umich.edu}. I thank Yuehao Bai, Marinho Bertanha, Steve Bond, Zach Brown, Adam Brzezinski, Jan De Loecker, Ying Fan, Florian Gunsilius, Andreas Joseph, Valentin Kecht, Tom Key, Harry Kleyer, Jagadeesh Sivadasan, Jonas Slaathaug Hansen, Dimitriy Stolyarov, Rick Van der Ploeg, Austin Wright, participants at the 2023 North America Econometric Society Meeting and the University of Michigan IO seminar, and an anonymous referee for the Bank of England Staff Working Papers for helpful discussion and comments. All errors are my own.}
}
\affil{University of Michigan}
\date{\today}

\begin{document}
\thispagestyle{empty}
\maketitle
\begin{abstract}
\noindent 
Production functions are potentially misspecified when revenue is used as a proxy for output. I formalize and strengthen this common knowledge by showing that neither the production function nor
Hicks-neutral productivity can be identified with such a revenue proxy. This result obtains when relaxing the standard assumptions used in the literature to allow for imperfect competition. It holds for a large class of production functions, including all commonly used parametric forms. Among the prevalent approaches to address this issue, only those that impose assumptions on the underlying demand system can possibly identify the production function.

\bigskip\noindent\textbf{JEL-Classification}:  L0, L11, C2, E23, D24.\\

\vfill
\end{abstract}

 \renewcommand{\thefootnote}{\arabic{footnote}}
 \pagebreak

\pagestyle{plain} \pagenumbering{arabic}

\section{Introduction}

The identification and estimation of production functions are fundamental problems in economics, essential for the study of productivity, returns to scale, elasticities of substitution, and markups. Production functions describe how firms produce physical goods from inputs, but the empirical estimation of these functions often relies on sales revenue (output times prices) as a proxy for quantity because revenue is typically observed while quantity is not. As prices are determined not only by production but also by demand-side factors, such an approach is likely to lead to biased estimates of the production function. This issue has long been recognized in the production function literature in various settings, though few formal results exist that precisely characterize what can and cannot be identified with revenue data. 

In this paper, I demonstrate formally that under the standard assumptions in the literature, neither the production function nor Hicks-neutral productivity can be identified from such a ``revenue production function'' when there is imperfect competition with heterogeneous prices. This result applies to a general class of ``weakly separable'' production functions, including all commonly used parametric forms such as the Cobb-Douglas and CES production functions. The result also extends to multi-product production functions when only product-level revenues, not outputs, are observed. Importantly, I show that the Markov assumption, which has traditionally been used to identify production functions, does not provide identification for revenue production functions. The assumptions involved require only cost minimization, not profit maximization, so the result holds for a wide range of market structures. Moreover, I establish that the only way to break the non-identification result in the presence of heterogeneous prices is by imposing demand-side restrictions. 

\subsection{Contribution}

\begin{figure}[htbp!]
\centering
\caption{Illustration of Non-Identification Mechanism}
\includegraphics[width = 0.7\textwidth]{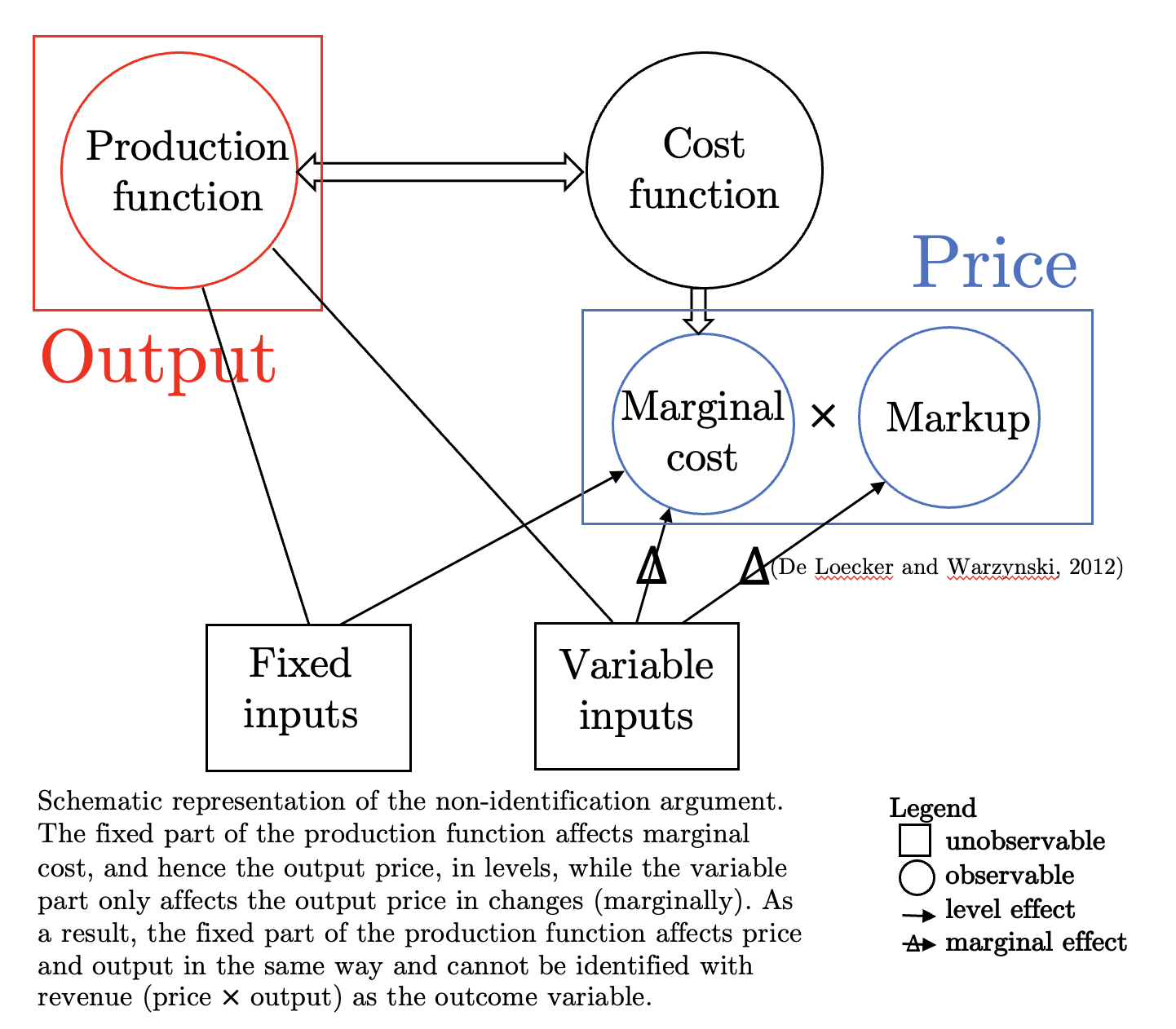}
\label{fig:diagram}
\end{figure}

The basic idea behind the non-identification result is illustrated in Figure \ref{fig:diagram}. The unobserved output price can be decomposed as the product of marginal cost and the markup. Both unobserved quantities are functions of the production and cost function, which are equivalent by duality \citep{shephard2015theory}. In particular, the fixed inputs affect these quantities through their effect on the production level, while the variable inputs affect them through their marginal effect on production. As a result, the fixed part of the production function affects output (the production level) and prices in the same way, and there is no identifying variation in a firm's revenue that allows one to back out the fixed inputs' effect on output.

The issue of unobserved output prices has long been recognized in the production function literature. Early papers such as \citet{abbott1992price, basu1997returns} and \citet{klette1996inconsistency} demonstrated that value-added and revenue production functions contain an additional term related to marginal cost or the markup. More recent papers that use datasets where output prices are observed have empirically confirmed these findings by documenting biases in estimates based on revenue production functions, compared to quantity production functions \citep{DeLoecker16, de2021hitchhiker, Mairesse2005panel, ornaghi2008price}, see \citet{DeLoecker2014firm} for a summary. Other papers have used approaches tailored to their empirical settings to address this bias without formally analyzing or quantifying it \citep{collard2015reallocation, smeets2013estimating, atalay2014materials, allcott2016electricity}, see \citet{deloecker2021comment} for further references.
This article aims to advance our understanding of this issue by formally analyzing how the revenue proxy leads to non-identification of not only the production function but also productivity. Based on these derivations, I can make precise statements about the sources of and potential solutions for this non-identification. \citet{Bond2020some} also consider the issue of identification, but they focus on the non-identification of the markup in a monopolistic competition setting. In contrast, this article establishes the non-identification of the entire production function and productivity under general conditions.

\subsection{Related Literature}
This article is relevant to several areas of literature. First, the literature on production function estimation and identification has developed a wide range of methods for estimating firm-level production functions under the assumption that physical output is observed and firms produce a single output \citep{Olley1996, Blundell2000gmm, Ackerberg15, Levinsohn03, Demirer2020production, Gandhi2017identification}. My results complement this literature by establishing precisely how one of the two most common deviations from these assumptions -- observed output prices -- leads to a failure of identification. Moreover, I show that this identification failure persists when relaxing the single-product assumption. 

This underscores the importance of the second strand of literature, which has focused on further extending production function estimation methods to the case when output \textit{is} observed, which raises issues related to multi-product production and the aggregability of production quantities \citep{deloecker2011product, DeLoecker16, dhyne2020theory,orr2019within, DeLoecker2014firm}. Since they establish strict limits on what can be done without observing price data, my results highlight the value of further developing these methods and the product-level production data they use.

Third, several strands of literature in industrial organization \citep{de2021industrial}, trade \citep{Keller2009multinational, amiti2007trade, bloom2016trade, brandt2017wto}, and international economics \citep{halpern2015imported} have aimed to estimate productivity \citep{syverson2011determines}, returns to scale, and more recently, markups \citep{DeLoecker12, DeLoecker19, Demirer2020production}, using production functions. In the absence of output data, many of these papers relied on a revenue proxy, to estimate these objects of interest. To do so, they developed a variety of approaches to address biases arising from the use of such a revenue proxy. My results formalize and generalize these approaches, and in particular, provide justification for the subset of approaches that relied on demand-side restrictions. 

Fourth, a strand of the macroeconomic literature, building on \citet{hall1989invariance}, has also dealt with the issue of unobserved output prices. Since this literature imposes strong assumptions on the production function, it can typically identify the production function \citep{basu1997returns}. Indeed, Corollaries \ref{corollary:cd} and \ref{prop:revenue_ces_basic} further below directly imply that under a constant returns to scale assumption combined with an assumption that the production function is of the Cobb-Douglas or Constant Elasticity of Substitution (CES) form, the production function can be fully identified. This combination of assumptions is prevalent in the macroeconomic production function literature but is restrictive in the context of the microeconomic literature, where estimating returns to scale is often one of the explicit goals. However, my results imply that, in the absence of such strong assumptions, the way this literature has tried to address the problem (using the growth rate of value added) does not lead to identification. In that case, one can also not identify the ``revenue productivity'' term studied in another strand of the macroeconomics literature \citep{foster2008reallocation,hsieh2009misallocation}.

\paragraph{\textbf{Notation.}} Vectors are bold, unknown variables are denoted by Greek letters, and observed variables are denoted by Latin letters. Levels of observed variables are capitalized, while natural logarithms are lowercase.


\section{Setting} \label{sec:setting}

I consider a general class of non-parametric production functions introduced by \citet{shephard1953cost}. The production function inputs are separated into dynamic inputs $\mathbf{X}_{it}$ and freely variable inputs $\mathbf{V}_{it}$. The vectors of associated input prices are denoted as $\mathbf{P}^X_{it}$ and $\mathbf{P}^V_{it}$. Hicks-neutral (total-factor) productivity is denoted as $\omega_{it}$, and the unobserved error term capturing ex-post output shocks and/or measurement error in output is denoted as $\varepsilon_{it}$. The information set of firm $i$ at time $t$, $\mathcal{F}_{it}$, includes $\omega_{it}$ and is assumed to be mean independent of ex-post output shocks. The ex-ante expected output shocks are denoted as $\mathcal{E}_{it} \coloneqq E[\exp(\varepsilon_{it}) | \mathcal{F}_{it}]$. The production technology is denoted as $F_t(\mathbf{X}_{it}, \mathbf{V}_{it})$, the planned (ex-ante) output as $Q^*_{it} = Q_t(\cdot) \exp(\omega_{it})$, and the observed (ex-post) output as $Q_{it} := Q^*_{it} 
\exp(\varepsilon_{it})$.

The production function literature imposes a few fundamental assumptions to identify the production function \citep{Olley1996, Levinsohn03, Ackerberg15}. I will impose these assumptions and show that when only revenue is observed, the production functions is not identifiable when relaxing one of the commonly imposed assumptions, that of perfect competition.

I start with the following general assumption on the production function,
\begin{assumption}[Weak homothetic separability] \label{asspt:homothetic}\
    \begin{enumerate}[(i)]
        \item The production function of firm $i$ at time $t$ is of the form
            \begin{equation} \label{eq:homothetic}
                Q_{it} = F_t\left( \mathbf{X}_{it}, h_t\left( \mathbf{X}_{it},   \mathbf{V}_{it} \right) \right)  \exp(\omega_{it}) \exp( \varepsilon_{it}), 
            \end{equation}
        \item with $h_{t}\left(\mathbf{X}_{i t}, \cdot \right) \text { homogeneous of arbitrary degree (homothetic) for all } \mathbf{X}_{it}$.
    \end{enumerate}
\end{assumption} 
\begin{assumption}[Cost minimization] \label{asspt:costmin} 
    \noindent The firm minimizes its short-term cost of production with respect to its freely variable non-separable inputs $\mathbf{V}_{it}$, given variable input prices $\mathbf{P}^V_{it}$ and the productivity shocks $\omega_{it}$.
\end{assumption}
Assumption \ref{asspt:homothetic} was introduced by \citet{shephard1953cost} and has recently been 
used by \citet{Demirer2020production} in the context of non-parametric identification of quantity production functions. It encompasses a large class of production functions, including all commonly used parametric ones \citep{Demirer2020production}. In particular, Assumption \ref{asspt:homothetic} (i) formulates a non-parametric production function and does not restrict the production function in the absence of further restrictions on $h(\cdot)$. Assumption \ref{asspt:homothetic} (ii) implies that ratio of the non-separable inputs' marginal product depends only on the inputs through their ratio. This assumption only restricts the class of models considered insofar as it requires that the inputs with respect to which the firm minimizes its short-term cost are homothetic in $h(\cdot)$. This does not mean that the entire production function has to satisfy homotheticity. Moreover, if only one input enters the firm's short-term cost, then (ii) can always be satisfied by rewriting the production function appropriately. For example, consider a standard three-input setup, with capital $K_{it}$, labor $L_{it}$, and material (or intermediate) inputs $M_{it}$. If labor $L_{it}$ is subject to some adjustment cost and therefore dynamically chosen, I have $\mathbf{V}_{it} = M_{it}$, and homotheticity of $h(\cdot)$ can always be satisfied by defining $h(\cdot)$ as the identity function. When both labor and material inputs are freely variable, however, the restriction on $h(\cdot)$ becomes meaningful.

Assumption \ref{asspt:costmin} is a commonly used assumption in the literature and allows for most forms of imperfect competition. It is therefore much weaker than the common assumption of perfect competition. Though the latter assumption implies that the unobserved prices are common across firms and can hence simply be absorbed by industry price deflators, it is very strong and rarely realistic in practice.

In addition, I require the following standard existence and regularity assumptions,
\begin{assumption}[Properties of the Production and Cost Functions] \label{asspt:properties_prodf} 
The production possibilities set satisfies the standard properties in Assumption \ref{asspt:properties_T} in Appendix such that the production function exists. Moreover, the production function satisfies,
\begin{enumerate}[(i)]
    \item If $h(\mathbf{X}_{it}, \mathbf{V}_{it}) > h(\mathbf{X}'_{it}, \mathbf{V}'_{it})$ then $F\left( \mathbf{X}_{it}, h\left(  \mathbf{X}_{it},  \mathbf{V}_{it} \right) \right) > F\left( \mathbf{X}_{it}, h\left(  \mathbf{X}'_{it},  \mathbf{V}'_{it} \right) \right)$ (strict monotonicity in variable inputs).
    \item The production function $F\left( \mathbf{X}_{it}, h\left(  \mathbf{X}_{it},  \mathbf{V}_{it} \right) \right)$ is continuous and differentiable in variable inputs $\mathbf{V}_{it}$ (duality). 
    \item The cost function $C\left(Q^*_{it}, \mathbf{X}_{it}, \mathbf{P}^V_{it}, \omega_{it}\right)$ is continuous and differentiable in planned output $Q^*_{it}$ (existence of the marginal cost function).
\end{enumerate}
\end{assumption}
Here, (i) slightly strengthens the weak monotonicity necessary for the existence of the production function, but is still weaker than strict monotonicity in all inputs, which is a standard assumption \citep[p.9]{chambers1988applied}. The assumption guarantees that the inverse of $F(\cdot)$, given $\mathbf{X}_{it}$, exists. Assumptions \ref{asspt:properties_prodf} (ii) and (iii) are standard differentiability assumptions \citep{chambers1988applied, mcfadden1978cost}. The continuity in (ii), combined with the assumptions on the production possibilities set, implies duality between the variable cost function and the production function \citep[Thm. 2]{diewert2022duality}. Part (iii) implies that the marginal cost function, $\dpar{C(\cdot)}{Q^*_{it}}$ exists. 

The following two assumptions are ubiquitous in the literature and have been instrumental to establishing identification of quantity production functions. Even though they are rather strong and as such should aid identification, I will show that they still fail to identify the production function in the context of revenue production functions. 

\begin{assumption}[Scalar Unobservable] \label{asspt:scalar}
Demand for (one of) the firm's inputs $v_{it}$ is given by,
\[ V_{it} = S_t(\mathbf{X}_{it}, V_{-it}, \omega_{it}), \]
where $V_{-it}$ denotes any freely variable input other than $V_{it}$ and input prices are assumed to be common across firms, $\mathbf{P}^V_{it}=\mathbf{P}^V_{t}$ such that they can be subsumed in $S_t$. 
\end{assumption}

This is simply a general way to write the scalar unobservable assumption used in most production function literature. \citet{Olley1996} first used this assumption for investment, not input demand. \citet{Levinsohn03} later proposed an alternative version for material input demand, and \citet{Ackerberg15} for conditional input demand. While some of these papers have imposed this assumption in an ad-hoc fashion, one can also formally derive it from cost minimization using Shephard's Lemma and assuming common input prices and homogeneous demand curves.\footnote{One can alternatively assume that input prices are observed, in which case the assumption would become $V_{it} = S_t(\mathbf{X}_{it}, V_{-it}, \omega_{it}, \mathbf{P}^V_{it})$ where the input prices now explicitly enter the function as an argument instead of being subsumed in the $t$ subscript on $S_t$.} For completeness, this is proved in Appendix \ref{sec_app:scalar}. If one instead derives the optimal input demand function from profit maximization, a price term or markup term will also show up in the expression (e.g. \citet[App.C.1]{Bond2020some}). This could give the false impression that the scalar unobservable assumption is inconsistent with the assumption that output prices are unobserved, or alternatively that they are a function of observables only.

These restrictive assumptions have played a crucial role in the identification arguments of these papers, by allowing one to substitute out for unobserved $\omega_{it}$ in the production function, predict target output $Q^*_{it}$, and form moments on the ex-post shock to productivity by leveraging the following assumption,
\begin{assumption}[Markov] \label{asspt:markov}
    Hicks-neutral productivity follows a first-order Markov process, 
    \begin{equation}
        \prob(\omega_{it} \mid \mathcal{F}_{it-1}) = \prob(\omega_{it} \mid \omega_{it-1}).
    \end{equation}
\end{assumption}
Together with the mean independence of ex-post output shocks, this assumption implies that one can write firms' Hicks-neutral productivity process as $
\omega_{it} = g(\omega_{it-1}) + \xi_{it}$,
where $E[\xi_{it} | \mathcal{F}_{it-1}] = 0$. Again, I emphasize that \textit{even} under these restrictive assumptions, the revenue production function cannot be identified. Relaxing them would only further reduce identifying information, so the non-identification result would continue to hold. Another way to phrase this is that non-identification arises from \textit{merely} relaxing the assumption of perfect competition used in the preceding literature.
\section{Non-Identification of Revenue Production Functions} \label{sec:main}

\subsection{Main Result}

 I am now equipped to prove the main result. To ease notation, I omit the time subscripts on all functions. Denote a firm's target revenue $R^*_{it} := P_{it}  Q^*_{it}$ where $P_{it}$ indicates firm-level output prices. Also, write $S^{*V}_{it} := \frac{P^V_{it}  V_{it}}{R^*_{it}}$ to indicate the revenue share of the input $V_{it} \in \mathbf{V}_{it}$. Then, we have 
\begin{theorem}[Non-Identification of the Production Function] \label{thm:nonIdentification_K} 

Let Assumptions \ref{asspt:properties_prodf}, \ref{asspt:homothetic}, \ref{asspt:costmin}, \ref{asspt:scalar}, and \ref{asspt:markov} hold. Furthermore, assume that production inputs $(\mathbf{X}_{it}, \mathbf{V}_{it})$, input prices $\mathbf{P}^X_{it}, \mathbf{P}^V_{it}$, and revenue $R_{it}$ are observed, but output prices $P_{it}$ and quantities $Q_{it}$ are not. Then the revenue production function can be written in terms of observables as,
\begin{equation}
    R_{it} =  G(\mathbf{X}_{it}, \mathbf{V}_{it}, \mathbf{P}^V_{it}, \mathbf{S}^{*\mathbf{V}}_{it})  \mathcal{E}_{it}^{-1} \exp(\varepsilon_{it}),
\end{equation}
and identification of $G(\cdot)$ is insufficient for identification of the production function.




\begin{proof}

The cost function for the class of production functions satisfying weak homothetic separability can be obtained from the general cost minimization problem, where the firm minimizes short-run costs given its level of planned output $Q^*_{it}$
\begin{equation} \label{eq:costmin}
\begin{aligned}
    & \min_{ \mathbf{V}_{it}}  \:  \mathbf{P}^V_{it} \cdot \mathbf{V}_{it} \\
    &  \text { s.t. }  E[ F\left( \mathbf{X}_{it}, h\left(  \mathbf{X}_{it},  \mathbf{V}_{it} \right) \right) \expo(\omega_{i t}) \exp(\varepsilon_{it}) | \mathcal{F}_{it} ]  \geqslant Q^*_{i t},
\end{aligned}
\end{equation}
Since $\mathcal{F}_{it}$ includes inputs and productivity shocks, I can rewrite the constraint as,
\begin{equation}
    F\left( \mathbf{X}_{it}, h\left(  \mathbf{X}_{it},  \mathbf{V}_{it} \right) \right) \expo(\omega_{i t})  \geqslant \tilde{Q}^*_{i t},
\end{equation}
with $\tilde{Q}^*_{i t} \coloneqq \frac{Q^*_{it}}{\mathcal{E}_{it}} \coloneqq \frac{Q^*_{it}}{E[\exp(\varepsilon_{it}) | \mathcal{F}_{it}]} $. Then, I can write the general cost function as \citep{chambers1988applied, Demirer2020production},
\begin{align*} \label{eq:demirer} 
    & C\left( \tilde{Q}^*_{it}, \mathbf{X}_{it}, \textbf{P}^V_{it}, \omega_{it} \right)  \\
    & =  \min_{ \mathbf{V}_{it}} \left\{ \mathbf{P}^V_{it} \cdot \mathbf{V}_{it} : 
    \tilde{Q}^*_{it} \leqslant  F\left( \mathbf{X}_{it}, h\left(  \mathbf{X}_{it},  \mathbf{V}_{it} \right) \right)  \expo(\omega_{i t})  \right\} \\
    & =  \min_{ \mathbf{V}_{it}} \left\{ \mathbf{P}^V_{it} \cdot \mathbf{V}_{it} : 
     F^{-1}\left( \mathbf{X}_{it}, \frac{\tilde{Q}^*_{it}}{\exp(\omega_{it})} \right)  \leqslant  h\left(  \mathbf{X}_{it},  \mathbf{V}_{it} \right) \right\} \\
    & =  \min_{ \mathbf{V}_{it}} \left\{ \mathbf{P}^V_{it} \cdot \mathbf{V}_{it} : 
    1 \leqslant  h\left(\mathbf{X}_{it}, \frac{ \mathbf{V}_{it} }{ F^{-1}\left( \mathbf{X}_{it}, \frac{\tilde{Q}^*_{it}}{\exp(\omega_{it})} \right)  } \right) \right\}  \\
    & =  \min_{ \mathbf{V}_{it}} \left\{  F^{-1}\left( \mathbf{X}_{it}, \frac{\tilde{Q}^*_{it}}{\exp(\omega_{it})} \right)(\mathbf{P}^V_{it} \cdot \mathbf{V}_{it}) : 
    1 \leqslant  h\left(  \mathbf{X}_{it},  \mathbf{V}_{it} \right) \right\} \\
    & =  F^{-1}\left( \mathbf{X}_{it}, \frac{\tilde{Q}^*_{it}}{\exp(\omega_{it})} \right)  \min_{ \mathbf{V}_{it}} \left\{ \mathbf{P}^V_{it} \cdot \mathbf{V}_{it} :
    1 \leqslant  h\left(  \mathbf{X}_{it},  \mathbf{V}_{it} \right) \right\}   \\ 
    & \coloneqq  F^{-1}\left( \mathbf{X}_{it}, \frac{\tilde{Q}^*_{it}}{\exp(\omega_{it})} \right)  C_2\left( \mathbf{X}_{it},  \mathbf{P}^V_{it} \right) \stepcounter{equation}\tag{\theequation} 
\end{align*}
where line 2 follows from Assumption \ref{asspt:properties_prodf} (strict monotonicity), line 3 and 4 follow from the assumption that $h(\cdot)$ is homothetic (since we can always redefine $F(\cdot)$ and $h(\cdot)$ to make $h(\cdot)$ homogeneous of degree one), and the last line defines a new function $C_2(\cdot)$.
To obtain the marginal cost, take the derivative of $C(\cdot)$ in Eq. \eqref{eq:demirer} with respect to target output, which exists by Assumption \ref{asspt:properties_prodf},
\begin{align*} 
\label{eq:marginal_cost}
    \lambda_{it} & = \frac{ \partial C\left( \tilde{Q}^*_{it}, \mathbf{X}_{it}, \textbf{P}^V_{it}, \omega_{it} \right) }{ \partial Q^*_{it} } \\
    & = \frac{ \partial F^{-1} }{ \partial x_2 }\left( \mathbf{X}_{it}, \frac{\tilde{Q}^*_{it}}{\exp(\omega_{it})} \right)
   \left(\expo(\omega_{it}) \mathcal{E}_{it} \right)^{-1}  
    C_2\left( \mathbf{X}_{it},  \mathbf{P}^V_{it} \right) \\ 
    & = \frac{ \partial F^{-1} }{ \partial x_2 } 
    \left(  \mathbf{X}_{it},  F\left( \mathbf{X}_{it}, h\left(  \mathbf{X}_{it},  \mathbf{V}_{it} \right) \right) 
    \right) 
    \left(\expo(\omega_{it}) \mathcal{E}_{it} \right)^{-1}  
    C_2\left( \mathbf{X}_{it},  \mathbf{P}^V_{it} \right)\\ 
    & = \frac{   C_2\left( \mathbf{X}_{it},  \mathbf{P}^V_{it} \right)
    }{ 
    \frac{ \partial F }{ \partial x_2 }  
    \left( \mathbf{X}_{it}, h\left(  \mathbf{X}_{it},  \mathbf{V}_{it} \right)  \right)
    \expo(\omega_{it}) \mathcal{E}_{it}
    }  \stepcounter{equation}\tag{\theequation} 
\end{align*}
where $\frac{\partial F(\cdot)}{\partial x_i}$ denotes the derivative of $F(\cdot)$ with respect to its ith argument, line 3 imposes that the constraint of the cost minimization problem is binding at the optimum such that $Q^*_{it} = F(\cdot)\expo(\omega_{it})$ and line 4 uses the inverse function theorem for the $\mathbb{R} \rightarrow \mathbb{R}$ function obtained when $\mathbf{X}_{it}$ is held fixed, for all values of $\mathbf{X}_{it}$. 

Finally, rewrite the output elasticity with respect to any of the (log) variable inputs $v_{it} \in \mathbf{v}_{it}$,
\begin{equation} \label{eq:output_elast} 
\begin{split}
    \frac{
     \partial f\left(\mathbf{X}_{it}, h\left(  \mathbf{X}_{it},  \mathbf{V}_{it} \right) \right)
     }{
     \partial v_{it}
     } 
     & = \frac{ 
     \partial F\left(\mathbf{X}_{it}, h\left(  \mathbf{X}_{it},  \mathbf{V}_{it} \right) \right)
     }{
     \partial  x_2
     }
     \frac{
     1
     }{
     F\left(\mathbf{X}_{it}, h\left(  \mathbf{X}_{it},  \mathbf{V}_{it} \right) \right)
     }
     \frac{
     \partial  h\left(  \mathbf{X}_{it},  \mathbf{V}_{it} \right)
     }{
     \partial v_{it}
     }, 
\end{split}
\end{equation}
where $f(\cdot)$ indicates the log production function.

Under imperfect competition, price equals markup, denoted as $\mu_{it}$, times marginal cost. Eq. \eqref{eq:marginal_cost} gives an expression for marginal cost. The markup can be written as the ratio of variable output elasticity to revenue share: $\mu_{it} = \frac{\partial f\left(\mathbf{X}_{it}, h\left(  \mathbf{X}_{it},  \mathbf{V}_{it} \right) \right) }{ \partial v_{it} } (S^{*V}_{it})^{-1}$ \citep{DeLoecker12, klette1999market}. I can plug this into the RHS of the production function in Eq. \eqref{eq:homothetic} and multiply the LHS by $P_{it}$ to get,
\begin{align*} \label{eq:revenuePF_nonparam_nonident} 
    R_{it} & = 
     F\left(\mathbf{X}_{it}, h\left(  \mathbf{X}_{it},  \mathbf{V}_{it} \right) \right) \expo(\omega_{it}) \lambda_{it} 
     \frac{\partial f\left(\mathbf{X}_{it}, h\left(  \mathbf{X}_{it},  \mathbf{V}_{it} \right) \right)
     }{
     \partial v_{it}
     } (S^{*V}_{it})^{-1} \exp(\varepsilon_{it}) \\
    & = 
    \frac{
         F\left(\mathbf{X}_{it}, h\left(  \mathbf{X}_{it},  \mathbf{V}_{it} \right) \right)\expo(\omega_{it}) C_2\left( \mathbf{X}_{it},  \mathbf{P}^V_{it} \right)\frac{ 
     \partial F\left(\mathbf{X}_{it}, h\left(  \mathbf{X}_{it},  \mathbf{V}_{it} \right) \right)
     }{
     \partial  x_2
     } \frac{\partial  h\left(  \mathbf{X}_{it},  \mathbf{V}_{it} \right)
     }{
     \partial v_{it}
     }
    }{ 
    \frac{\partial F\left(\mathbf{X}_{it}, h\left(  \mathbf{X}_{it},  \mathbf{V}_{it} \right) \right)
     }{
     \partial  x_2
     } \expo(\omega_{it}) \mathcal{E}_{it} S^{*V}_{it} F\left(\mathbf{X}_{it}, h\left(  \mathbf{X}_{it},  \mathbf{V}_{it} \right) \right)
    } \exp(\varepsilon_{it})  \\ 
     & = 
       C_2\left( \mathbf{X}_{it},  \mathbf{P}^V_{it} \right)   \frac{\partial  h\left(  \mathbf{X}_{it},  \mathbf{V}_{it} \right)
     }{
     \partial v_{it}
     } (S^{*V}_{it} \mathcal{E}_{it})^{-1} \exp(\varepsilon_{it}) \stepcounter{equation}\tag{\theequation}
\end{align*}
which gives $M$ equations for $R_{it}$, one for each flexible input $V$. 

Thus, I obtain an expression for the revenue production function in terms of observables. To establish that the production function $Q(\cdot)$ cannot be identified from this equation, assume without loss of generality that $\dpar{h(\cdot)}{v_{it}}$ and $C_2(\cdot)$ can be identified from the last line of Eq. \eqref{eq:revenuePF_nonparam_nonident}.\footnote{This is without loss of generality as it is the best-case identification of the functions involved.} Then $h(\cdot)$ is identified up to a constant. Now, since $h(\cdot)$ is a component of the composite function $Q(\cdot)$, it is intuitive that identification of $h(\cdot)$ alone is not sufficient for identification of $Q(\cdot)$. This is proved formally in Lemma \ref{lemma:composite} in Appendix \ref{sec_app:aux} for completeness. Moreover, since the cost function was shown to equal  $F^{-1}\left( \mathbf{X}_{it}, \frac{\tilde{Q}^*_{it}}{\exp(\omega_{it})} \right)  C_2\left( \mathbf{X}_{it},  \mathbf{P}^V_{it} \right) $ in Eq. \eqref{eq:demirer}, the proof of Lemma \ref{lemma:composite} also immediately implies that identification of $C_2(\cdot)$ is not sufficient for identification of the cost function.\footnote{This follows because it is shown in Lemma \ref{lemma:composite} that the class of functions $F$, $\mathcal{F}$ that satisfy the stated assumptions is not a singleton set, and hence the class of inverse functions $F^{-1}$ is neither.} But by Shephard's Duality Theorem \citep[Thm. 2]{diewert2022duality}, the cost function is dual to the production function.\footnote{That is, the cost function is necessary and sufficient for the production function, under the assumption of cost minimization and firms being input price-takers. This can also be seen from Eq. \eqref{eq:demirer}} As a result, identification of $C_2(\cdot)$ is also not sufficient for identification of $Q(\cdot)$. Furthermore, it is clear that no composition of $C_2(\cdot)$ and $h(\cdot)$ can identify $Q(\cdot)$ either, since obtaining $Q(\cdot)$ from a composition of either function requires that the composing functions equal either $F(\cdot)$ or $F^{-1}(\cdot)$, which by the preceding arguments clearly do not coincide with $C_2(\cdot)$ or $h(\cdot)$, respectively. 

Finally, note that combining the $M$ versions of Eq. \eqref{eq:revenuePF_nonparam_nonident} gives,
\begin{align*}
 R_{it}&  =   C_2\left( \mathbf{X}_{it},  \mathbf{P}^V_{it} \right)   \left( \frac1M \sum_{V \in \mathbf{V}}  \frac{\partial  h\left(  \mathbf{X}_{it},  \mathbf{V}_{it} \right)
     }{
     \partial \log V_{it}
     } (S^{*V}_{it} )^{-1} \right) \mathcal{E}_{it}^{-1} \exp(\varepsilon_{it}) \\
     & \coloneqq G(\mathbf{X}_{it}, \mathbf{V}_{it}, \mathbf{P}^V_{it}, \mathbf{S}^{*\mathbf{V}}_{it})  \mathcal{E}_{it}^{-1}  \exp(\varepsilon_{it}). \stepcounter{equation}\tag{\theequation}
\end{align*}
By the preceding arguments about $C_1(\cdot), h(\cdot)$, clearly, identification of $G(\cdot)$ is in general also not sufficient to identify $Q(\cdot)$. Moreover, $G(\cdot)$ takes all observable variables (prices, inputs, and revenue shares) as its arguments. Hence, there is no alternative way of rewriting $G(\cdot)$ as some function $H(\mathbf{X}_{it}, \mathbf{V}_{it}, \mathbf{P}^V_{it}, \mathbf{S}^{*\mathbf{V}}_{it})$ using production-side model equations such that identification of $H(\cdot)$ would be sufficient for identification of $Q(\cdot)$, as any such $H(\cdot)$ would have to be a composite function of $G(\cdot)$.  

Finally, I need to argue that Assumption \ref{asspt:markov} cannot deliver identification of $Q(\cdot)$ either. This is intuitive since the Markov assumption was introduced in the literature to handle the unobserved productivity term $\omega_{it}$ in the quantity production function -- but this term drops out in the revenue production function. Indeed, the way the preceding literature has used this assumption for identification is by rewriting $\omega_{it}$ as a function of model parameters and ``observables'',
\begin{equation} \label{eq:omega_identification}
\omega_{it} = q^*_{it-1} - q(\mathbf{X}_{it-1}, \mathbf{V}_{it-1})
\end{equation}
where (log) target output $q^*_{it} = q(\mathbf{X}_{it}, \mathbf{V}_{it}) + \omega_{it}$ would be identified (and thus ``observed'') by substituting out $\omega_{it}$ in the (quantity) production function using the scalar unobservable Assumption \ref{asspt:scalar} and projecting out the ex-post output shock,
\begin{equation} \label{eq:q_star}
q_{it} = q^*(\mathbf{X}_{it}, \mathbf{V}_{it}) + \varepsilon_{it}.
\end{equation}
Then, identification would obtain from forming moments on the ex-post shock to $\omega_{it}$, $\xi_{it}$ \citep{Ackerberg15,Gandhi2017identification},
\begin{equation} \label{eq:markov_moment}
E\big[ \xi_{it} | \mathcal{F}_{it-1}\big] = E\big[ (q^*_{it} - q(\mathbf{X}_{it}, \mathbf{V}_{it})) - g\left( q^*_{it-1} - q(\mathbf{X}_{it-1}, \mathbf{V}_{it-1}) \right) | \mathcal{F}_{it-1} \big] = 0 
\end{equation}
Since a firm's output $Q_{it}$ is, however, not observed, the target output function $Q^*(\cdot)$ cannot be identified through the usual conditional moment condition 
for Eq. \eqref{eq:q_star}, $E[\varepsilon_{it} | \mathcal{F}_{it-1}] = 0$. Even if one could identify target revenue \[R^*_{it} = P_{it}  Q^*_{it} = G(\mathbf{X}_{it}, \mathbf{V}_{it}, \mathbf{P}^V_{it}, \mathbf{S}^{*\mathbf{V}}_{it}) \mathcal{E}_{it}^{-1} \] from the revenue production function,\footnote{For example, a full independence assumption makes $E[\mathcal{E}_{it} | \mathcal{F}_{it}]$ a constant \citep[p.2979]{Gandhi2017identification}.} one could still not identify $Q^*_{it}$ from this as $P_{it}$ is, of course, unobserved by assumption. Moreover, since I have shown that $Q(\cdot)$ cannot be identified from the revenue production function, and since $Q^*_{it} = Q(\cdot) \exp(\omega_{it})$, I conclude that $Q^*_{it}$ cannot be identified from the revenue production function. Hence, Eq. \eqref{eq:markov_moment} depends on an unobservable, $q^*_{it}$, and cannot deliver identification of $q(\cdot)$ either.
\end{proof}

\end{theorem}

I immediately obtain the following corollaries. 

\begin{corollary}[Non-Identification From Markov] \label{thm:nonIdentification_markov}
     The Markov Assumption \ref{asspt:markov} does not lead to identification of the production function when output quantities are unobserved.
\end{corollary}

This corollary is embedded in the main theorem but is worth emphasizing as the Markov assumption has been the main identifying assumption in the production function literature thus far. 

\begin{corollary}[Non-Identification From First-Order Conditions] \label{corr:foc}
The cost minimization first-order conditions can be rewritten in terms of observables as, 
 \label{thm:foc}
\begin{equation}  \label{eq:foc}
     P^V_{it} = C_2\left( \mathbf{X}_{it},  \mathbf{P}^V_{it} \right) \frac{ \partial h( \mathbf{X}_{it}, \mathbf{V}_{it} ) }{ \partial V_{it} }  \mathcal{E}_{it}^{-1} \quad \forall \: V \in \mathbf{V},
\end{equation}
and hence contain no additional information relative to the revenue production function.
 \begin{proof}
Impose equality on the constraint in Eq. \eqref{eq:costmin} by the strict monotonicity assumption of the production function with respect to $\mathbf{V}_{it}$, take the first derivative, and plug in the expression for $\lambda_{it}$ from Eq. \eqref{eq:marginal_cost}. This gives Eq. \eqref{eq:foc}, which is implied by the revenue production function.
\end{proof}
\end{corollary}

 Thus, the first-order conditions depend on the same functions as the revenue production function does. As a result, neither can they be used to identify $F(\cdot)$. This is, of course, not surprising as we used these first-order conditions to derive the result in Theorem \ref{thm:nonIdentification_K}. Yet, it is also worth emphasizing, given that many papers have used first-order conditions for the estimation of production functions, albeit under additional conditions that together do provide identification \citep[VI.B]{Gandhi2017identification}.

\begin{corollary}[Non-Identification of Output Elasticities] \label{thm:nonIdentification_elasticities}
     The short-run and long-run output elasticities $ \frac{
     \partial f\left(\mathbf{X}_{it}, h\left(  \mathbf{X}_{it},  \mathbf{V}_{it} \right) \right)
     }{
     \partial v_{it}
     } ,   \frac{
     \partial f\left(\mathbf{X}_{it}, h\left(  \mathbf{X}_{it},  \mathbf{V}_{it} \right) \right)
     }{
     \partial x_{it}
     }$ cannot be identified from a revenue production function.
\end{corollary}

This corollary follows directly from the non-identification of $F(\cdot)$ proved in Theorem \ref{thm:nonIdentification_K} together with Lemma \ref{lemma:composite}. Nonetheless, it is important to emphasize, as it highlights the contrast between this article's non-identification result and related results proved in \citet{Gandhi2017identification, Demirer2020production}. In particular, while those papers establish, under some additional conditions, the non-parametric non-identification of the production function when prices \textit{are} observed, they also prove that at least the flexible output elasticities are identified. By contrast, in my setting with unobserved output prices, neither the production function nor the output elasticities are identified. Moreover, note that the non-identification results in those papers rest on additional assumptions. In particular, \citet{Demirer2020production} considers a production function with both Hicks-neutral and labor-augmenting productivity, while \citet{Gandhi2017identification} assume that output prices are common across firms. Indeed, the non-identification result in \citet[Thm.1]{Gandhi2017identification} assumes that there is no cross-sectional or time-series variation in prices. When output prices are observed and vary across firms and time, the production function could, in fact, be identified fully using the proxy variable approach in \citet{Ackerberg15}. The same applies to the setting in \citet{Demirer2020production} when there is only Hicks-neutral productivity. By contrast, my result implies that absent observed output prices, neither the production function, nor the output elasticities, nor Hicks-neutral productivity (discussed below), can be identified. As such, this is a much stronger non-identification result. 
Finally, an insight used in the recent literature to provide identification is that the ratio of revenue shares equals the ratio of output elasticities \citep{Demirer2020production, Doraszelski2019using, Gandhi2017identification},
\[ \frac{S^V_{it}}{S^{V-}_{it}} = 
\frac{\frac{
     \partial f\left(\mathbf{X}_{it}, h\left(  \mathbf{X}_{it},  \mathbf{V}_{it} \right) \right)
     }{
     \partial v_{it}
     } }{\frac{
     \partial f\left(\mathbf{X}_{it}, h\left(  \mathbf{X}_{it},  \mathbf{V}_{it} \right) \right)
     }{
     \partial v_{-it}
     } }.
\]
However, this ratio does not provide any additional identifying information in a revenue production function setting. To see this, note that by \eqref{eq:output_elast}, this ratio equals $\nicefrac{\frac{
     \partial  h\left(  \mathbf{X}_{it},  \mathbf{V}_{it} \right)
     }{
     \partial v_{it}
     }}{\frac{
     \partial  h\left(  \mathbf{X}_{it},  \mathbf{V}_{it} \right)
     }{
     \partial v_{-it}
     }}$
which provides no additional information that could be used to identify $F(\cdot)$.

\begin{corollary}[Non-Identification of Productivity] \label{thm:nonIdentification_omega}
     Hicks-neutral productivity cannot be identified from a revenue production function.
\end{corollary}

Again, this corollary follows directly from the main theorem, but is worth emphasizing given the fact that production function estimation has been widely used for the estimation of productivity \citep{de2021industrial}. 
 The intuition behind this corollary is that marginal cost -- and, hence, output price -- depends inversely and one-to-one on Hicks-neutral productivity, since the latter linearly scales the production function. That is, an increase in Hicks-neutral productivity by construction increases output one-to-one. As a result, it also decreases marginal cost one-to-one, and these two effects cancel out in the revenue production function.

\subsubsection{Discussion}

Having established the main result, I 
now discuss two open questions: what does non-identification mean in this context, and how could it be resolved? 

There is a simple intuition behind Theorem \ref{thm:nonIdentification_K}. Under the assumption of variable cost minimization, and the assumption of price-setting implicit in the markup formula (markup equals price divided by marginal cost), the dynamic inputs $\mathbf{X}_{it}$ will only affect a firm's price through the \textit{level} of the production function, while the variable inputs $\mathbf{V}_{it}$ also affect the price through the gradient of the production function. The idea here is that small changes in variable inputs should affect a firm's price because both variable inputs and price can be adjusted instantaneously under the given assumptions. Changes in the dynamic inputs, however, only affect overall, not marginal production capacity, and hence only affect prices in that way as well. As a result, if I re-express output prices in terms of production-side variables, the part of the production function associated with the dynamic inputs, which is assumed to be separable, exactly cancels out, because the effect of these dynamic inputs on output feeds through one-to-one to output prices. 

The driver of this mechanism is that I can express output prices in terms of only production-side observables and revenue shares. This is the insight that motivated the so-called ``production approach'' to markup estimation, which relies on the fact that the markup can be expressed as the ratio of output elasticity to revenue shares \citep{deloecker2011product, klette1999market}. The intuition behind this fact rests on a simple cost-benefit condition. The output elasticity tells us the percent increase in output -- and thus, for fixed prices, revenue -- for a percent increase in one of the inputs. The revenue share, on the other hand, tells us what percent of revenue is spent on input costs for each percent increase in one of the inputs (since $\frac{\partial  s^{*V}_{it}}{\partial m_{it}} = S^{*V}_{it}$). As long as the output elasticity is larger than the revenue share, the firm is making a profit. In fact, the gap between both is exactly equal to the markup.

The literature has proposed several solutions to the lack of data on output prices. First, as is well-known, the use of industry deflators cannot resolve the issue when there is cross-sectional output price variation, so this approach only supports a highly restrictive set of models of imperfect competition \citep{deloecker2021comment} or perfect competition. Indeed, it only supports models of competition that imply homogenous prices, in which case industry-level price deflators equal firm-level prices up to some constant. 

Second, the introduction of input prices on the right-hand side of the revenue production function can also not alleviate the issue \citep{DeLoecker2014firm, DeLoecker19, deloecker2021comment}. This follows by the same logic as Corollary \ref{corr:foc}, as Eq. \eqref{eq:foc} gives an expression for input prices that depends only on the components of the revenue production function. This shows that input prices cannot introduce any additional identifying information into the revenue production function. 

This leaves, third, the imposition of assumptions on the underlying demand system. When a parametric demand system is specified, this approach can provide an explicit expression for output prices that does not depend directly on the production function \citep{deloecker2011product, Gandhi2017identification}. In some cases, non-parametric restrictions on the demand system may alternatively require an exact set of additional demand-side variables to be included in the model \citep{DeLoecker16}. In general, insofar as Theorem \ref{thm:nonIdentification_K} establishes non-identification under a general non-parametric revenue production function that depends on all production-side observables, it is clear that identification can only be established by introducing additional non-production (demand-side) variables. This requires restrictions on the underlying demand system. Of course, this by itself does not guarantee identification -- that will need to be established in the context of the particular assumptions imposed. Finally, the growth rate of revenue \citep{basu1997returns} provides no further information about the production function. This follows directly from the fact that the revenue \textit{level} does not identify the production function in a single period, so neither will it do so across periods.

The importance of these non-identification results is twofold. First, most parametric production functions satisfy weak homothetic separability, which implies that previous studies that estimated productivity, returns to scale, or markups using revenue as a proxy for output without imposing further structure are likely unidentified. To be clear, this does not mean that the methods used in these papers cannot deliver identification when output \textit{is} observed, or when some demand-side restrictions are imposed. Second, since the preceding results apply to a general class of production functions and market structures, and demonstrate not just the misspecification but the non-identification of revenue production functions, they underscore the importance of developing suitable estimators and datasets that address these issues \citep{dhyne2020theory}. Finally, in the absence of explicit consideration of this non-identification, these results suggest that practitioners should avoid estimating production functions and productivity with revenue data. Below, I further illustrate the results in light of the two most commonly used parametric functional forms.

\subsection{What Can Be Identified In Leading Parametric Cases?}

Next, I demonstrate Theorem \ref{thm:nonIdentification_K} for two leading parametric production functions, the Cobb-Douglas and CES production function. Since this requires a parametrization, I let $\mathbf{X}_{it} = K_{it}, \mathbf{V}_{it} = (L_{it}, M_{it})$, but the non-identification does not depend on these particular choices, as long as there is at least one dynamic input. 

\subsubsection{Cobb-Douglas}

For the Cobb-Douglas production function,
\begin{equation} \label{eq:CD_levels}
    Q_{it} = K_{it}^{\beta_K} \ L_{it}^{\beta_L} M_{it}^{\beta_M} \text{exp}(\omega_{it})\expo(\varepsilon_{it}),
\end{equation}
the (log) revenue production function can be written as,
\begin{equation} \label{eq:CD_revenue}
\begin{split}
    r_{it} = 
    \theta_0 +  {\frac{\beta_L}{\beta_L + \beta_M}}  (l_{it} + p^L_{it}) + {\frac{\beta_M}{\beta_L + \beta_M}}  (m_{it} + p^M_{it})
     - s^{*V}_{it} - \log \mathcal{E}_{it} +
    \varepsilon_{it}; \\ 
\end{split}
\end{equation}
where  $ \theta_0 = -\log(\beta_L + \beta_M) + \frac{\beta_M - \beta_L}{\beta_L + \beta_M}\left( \log\beta_L -\log\beta_M \right) + \log \beta_V$ for some $V \in (L, M)$. See Appendix \ref{sec_app:derivations} for the full derivations.
This equation illustrates Theorem \ref{thm:nonIdentification_K}, as the coefficient on capital, which determines the non-separable part of the production function, is not identified. Moreover, identification of the separable part, $h(L_{it}, M_{it}) = L_{it}^{\frac{\beta_L}{\beta_L+\beta_M}} M_{it}^\frac{\beta_M}{\beta_L + \beta_M}$, only allows me to identify the ratio of short-run output elasticities. Corollary \ref{thm:nonIdentification_omega} is also illustrated since $\omega_{it}$ does not show up in these expressions. I obtain the following,

\begin{corollary}[Non-Identification of the Revenue Cobb-Douglas] \ \label{corollary:cd}
The revenue Cobb-Douglas production function \eqref{eq:CD_revenue} only identifies the ratio of variable output elasticities, $\beta_L, \beta_M$. Neither the dynamic output elasticity $\beta_K$ nor productivity $\omega_{it}$ are identified. 

\end{corollary}

\subsubsection{CES Production Function}

Write the CES production function as,
\begin{equation}
Q_{it} = \left( (1-\beta_L - \beta_M) K_{it}^\sigma + \beta_L  L_{it}^\sigma + \beta_M M_{it}^\sigma \right)^{\frac{v}{\sigma}} \exp(\omega_{it}) \exp(\varepsilon_{it}),
\end{equation}
where $\beta_L, \beta_M$ are the share parameters, $v$ is the return to scale parameter, and $\sigma$ is the elasticity of substitution parameter. The (log) revenue equivalent is,
\begin{equation} \label{eq:CS_revenue}
    r_{it} =  \log \beta_V  + \sigma v_{it} + \frac{1-\sigma}{\sigma} \log \left( \beta_L  L_{it}^\sigma + \beta_M M_{it}^\sigma  \right) +   \frac{\sigma-1}{\sigma} \log B - s_{it}^{*V} - \log \mathcal{E}_{it} + \varepsilon_{it},
\end{equation}
where the unit cost function $B := (P^L_{it})^{\frac{\sigma}{\sigma-1}} \beta_L^{-\frac{1}{\sigma-1}} + (P^M_{it})^{\frac{\sigma}{\sigma-1}} \beta_M^{-\frac{1}{\sigma-1}} $. Again, following Theorem \ref{thm:nonIdentification_K}, one can see that the non-separable part of the production function is not identified, as the returns to scale $v$ do not show up in this expression. Similarly, $\omega_{it}$ again drops out, in line with Corollary \ref{thm:nonIdentification_omega}. The below proposition establishes what can be identified from a (potentially nested) revenue CES production function,

\begin{corollary}[Non-Identification of the Revenue CES] \label{prop:revenue_ces_basic}
    The revenue CES production function (Eq. \eqref{eq:CS_revenue}) can at most identify the elasticity of substitution $\sigma$ and the\textit{ ratio }of short-run distribution parameters $\frac{\beta_L}{\beta_M}$. Neither the returns to scale $v$, the output elasticities, nor Hicks-neutral productivity $\omega_{it}$ are identified. \\
    \textbf{Proof.} See Appendix \ref{sec_app:proofs}.
\end{corollary}

\subsection{Extension to Multi-Product Production Functions}

So far, I have maintained the assumption that firms manufacture a single product, which has been standard in the production function literature. However, recent studies have begun to explore the identification and estimation of multi-product production functions when prices are observed \citep{deloecker2011product, DeLoecker16, orr2019within, dhyne2020theory, valmari2023}. The present article addresses a distinct question: what can be identified in the absence of observed prices? Nonetheless, there is an intriguing extension of this question to the multi-product case: what can be identified when prices are unobserved, yet product-level sales data are available? These types of data are prevalent in the trade and international economics literature, as firm-to-firm transaction and export datasets often record product-level sales but not outputs \citep{munch2014decomposing, bernard2022origins}. In this section, I extend the non-identification result in Theorem \ref{thm:nonIdentification_K} to show that the multi-product production function can also not be identified when only product-level revenue is observed. 

There are two main approaches to multi-product production: separable and joint. 
Separable production assumes that firms produce their products independently so that the multi-product production function is just a bundle of single-product functions for each firm \citep{deloecker2011product,  DeLoecker16, orr2019within, valmari2023}. Under separable production, the non-identification result in Theorem \ref{thm:nonIdentification_K} applies directly, as in that case one can just replace the firm index $i$ by a firm-product index $ij$. In particular, under the stated assumptions the product-level production function would take the form, 
 \begin{equation} \label{eq:homothetic_multi}
                Q_{ijt} = F_{jt}\left( \mathbf{X}_{ijt}, h_{jt}\left( \mathbf{X}_{ijt},   \mathbf{V}_{ijt} \right) \right)  \exp(\omega_{ijt}) \exp( \varepsilon_{ijt}), 
            \end{equation}
and, following exactly the same derivation as for the proof of Theorem \ref{thm:nonIdentification_K}, the product-level revenue production function becomes,
\begin{equation}
    R_{ijt} =  G_j(\mathbf{X}_{ijt}, \mathbf{V}_{ijt}, \mathbf{P}^V_{ijt}, \mathbf{S}^{*\mathbf{V}}_{ijt})  \mathcal{E}_{ijt}^{-1} \exp(\varepsilon_{ijt}).
\end{equation}
By the same arguments as above, identification of $G_j$ in this case would not be sufficient for identification of $F_j$. Moreover, this identification of $G_j$ is not self-evident either, since it requires a way to back out product-level inputs $\mathbf{X}_{ijt}, \mathbf{V}_{ijt}$, which are rarely observed \citep{DeLoecker16, orr2019within, valmari2023}. 

Finally, an assumption that is commonly imposed in the separable approach in order to back out or aggregate the unobserved product-level inputs is that the technology $F_j$ is common across products, that is, there is no index $j$ on the production technology $F$ \citep{deloecker2011product, orr2019within}. However, the non-identification result holds even when assuming that one observes the product-level inputs and \textit{can} identify $G_j$, so this assumption clearly does not break it.

Joint production, on the other hand, models the production of products within a firm as being interdependent. Denote $Q_{ijt}$ the output of product $j$ by firm $i$ at time $t$. Moreover, denote $\mathbf{Q}_{-ijt} \coloneqq (Q_{i1t}, \ldots, Q_{i(j-1)t}, Q_{i(j+1)t},  \ldots, Q_{iJt})$ the output vector of the $J-1$ products other than $j$ produced by firm $i$.  Assuming that the multi-product transformation function satisfies the conditions in \citet{dhyne2020theory} so that the production function is well-defined, I consider the following class of production functions,

\begin{assumptionp}{\ref{asspt:homothetic}$'$}[Weak homothetic separability]\ 
    \begin{enumerate}[(i)]  
    \item The joint multi-product production function is of the form, \begin{equation}
        Q_{ijt} = F_{jt}\left(\mathbf{Q}^*_{-ijt} ,  \mathbf{X}_{it}, h_{jt}\left(\mathbf{Q}^*_{-ijt} ,  \mathbf{X}_{it},  \mathbf{V}_{it} \right)\right) \expo(\omega_{ijt})  \exp(\varepsilon_{ijt}),
    \end{equation}
    \item \noindent with $h_{jt}\left(\mathbf{Q}^*_{-ijt} ,  \mathbf{X}_{it},  \mathbf{V}_{it} \right)$ homogeneous of arbitrary degree for any $(\mathbf{Q}^*_{-ijt},\mathbf{X}_{it})$. \\
    \end{enumerate}
\end{assumptionp}
The interpretation of this assumption is identical to Assumption \ref{asspt:homothetic}: firms minimize their short-term variable cost with respect to the ratio of variable inputs $\mathbf{V}_{it}$. The only differences are that 1) these variable inputs are now used to jointly produce an entire product gamut and 2) the amount a firm produces of one product is jointly determined by its production -- and hence its marginal cost -- of the other products. 
Assumption \ref{asspt:costmin} equally remains nearly unchanged, except that the productivity term $\omega_{ijt}$ is now product-specific. In particular, firms solve the following cost minimization problem \citep{mcfadden1978cost, dhyne2020theory},
\begin{align*}
     &  \min_{  \mathbf{V}_{it}} \mathbf{P}^V_{it} \cdot \mathbf{V}_{it} \\
 &   \text{ s.t. }   E\big[ F_{j}\left(\mathbf{Q}^*_{-ijt} ,  \mathbf{X}_{it}, h_j\left(\mathbf{Q}^*_{-ijt} ,  \mathbf{X}_{it},  \mathbf{V}_{it} \right)\right) \expo(\omega_{ijt})  \exp(\varepsilon_{ijt}) \big| \mathcal{F}_{it} \big] \geq Q^*_{ijt},
 \end{align*}
where the variables denote the same objects as before, but with different indices. Since inputs and productivity are in the firm's information set $\mathcal{F}_{it}$, the vector of target outputs $\mathbf{Q}^*_{-ijt}$ is as well, and following the same steps as in the proof of Theorem \ref{thm:nonIdentification_K}, I get a product-specific variable cost function of the form,
\begin{equation}
 C_j\left(\tilde{Q}^*_{ijt}, \mathbf{Q}^*_{-ijt}, \mathbf{X}_{it},  \mathbf{P}^V_{it}, \omega_{ijt} \right),
\end{equation}
with the only difference with the single-product cost function being that the cost of one product depends on the production levels of all the other products as well. With this in hand, the rest of the proof follows analogously by carrying the additional argument $\mathbf{Q}^*_{-ijt}$. The resulting product-level revenue production function is, 
\begin{equation} \label{eq:revenueProdF_noPF}
\begin{split} 
    R_{ijt} & = 
       C_{j,2}\left(\mathbf{Q}^*_{-ijt}, \mathbf{X}_{it},  \mathbf{P}^V_{it} \right) \frac{ \partial h_j( \mathbf{Q}^*_{-ijt}, \mathbf{X}_{it}, \mathbf{V}_{it} ) }{ \partial \log V_{it} }
       (S^{*V}_{ijt} \mathcal{E}_{ijt})^{-1} \exp(\varepsilon_{ijt}),
\end{split}
\end{equation}
where $R^*_{ijt} = P_{ijt} Q^*_{ijt}$ and $S^{*X}_{ijt} = \frac{ P^X_{it} X_{it} }{ R_{ijt} }$ for any variable input $X_{it}$. This has an intuitive interpretation: the revenue a firm makes from one product depends on how much it produces of its other products. Thus, the revenue production function now depends on an additional vector of unobservables, $\mathbf{Q}^*_{-ijt}$, and $G_j(\cdot)$ cannot be identified. 

\section{Conclusion} \label{sec:conclusion}

This article has shown that neither the production function nor Hicks-neutral productivity can be identified when using revenue as a proxy for output. This holds true for a general class of production functions and under standard assumptions. There is no way to rewrite the revenue production function in terms of observable production-side variables to break this non-identification. The only approach that can obtain identification in this scenario is to impose restrictions on the underlying demand system. The non-identification was demonstrated parametrically for the Cobb-Douglas and CES production functions. It was also shown to hold for both separable and joint multi-product production functions. This work formalizes and generalizes previous findings that revenue production functions are misspecified and generate biased estimates.

The implications are significant. In the absence of observed output prices, perfect competition, or further restrictions on demand, practitioners should avoid estimating production functions using revenue as a proxy for output. Most objects of interest such as productivity, returns to scale, and markups cannot be identified. However, markups may still be estimated without production functions using the production approach to markup estimation, subject to additional restrictions \citep{deloecker2021comment}.

Future research may focus on establishing partial identification of the production function when output prices are unobserved \citep{flynn2019measuring}, or on further developing production function theory for when they \textit{are} observed \citep{dhyne2020theory, DeLoecker16}.

\clearpage 
\addcontentsline{toc}{chapter}{References}
\bibliographystyle{plainnat}
\bibliography{References}

\appendix

\renewcommand{\theequation}{\thesection.\arabic{equation}}

\clearpage 

\section{Auxiliary Results} \label{sec_app:aux}

\begin{lemma} \label{lemma:composite}
Let $h: W  \to Y_1, F: X \times Y_1 \to Z, Q^* : (w, y_2) \in W \times Y_2 \to F(y_2, h(w))$, with $W \coloneqq (X, V)$ for some $F \in \mathcal{F}, h \in \mathcal{H}, Q^* \in \mathcal{Q}$ where $\mathcal{F}, \mathcal{H}, \mathcal{Q}$ are spaces of composite and composing functions for which Assumptions \ref{asspt:properties_prodf} and \ref{asspt:homothetic} are satisfied. Then, for a given $h \in \mathcal{H}$ and a corresponding $q \in \mathcal{Q}, f \in \mathcal{F}$ such that $q= f(x, h(w))$ for all $w \in W, x \in X$; if $h$ is given but $f$ not, then $q$ can in general not be uniquely determined.
\begin{proof}
Suppose not. Then for any $\tilde{h} \in \mathcal{H}$, $\exists \tilde{f} \in \mathcal{F}$ such that, $\forall Q^* \in \mathcal{Q}, Q^* = \tilde{f}(y_2, \tilde{h}(x)),$ $\forall w \in W, x \in X$. Since $\mathcal{F}$ is defined as the set of component functions which, composed with some function in $\mathcal{H}$, give a function in $\mathcal{Q}$, this means that $\mathcal{F}$ must be a singleton set of functions $ \{ \tilde{f} \}$. In general, it is clear that the set of functions $\mathcal{F}$ that satisfy Assumption \ref{asspt:homothetic} and Proposition \ref{prop:properties_F} is not a singleton. 

For example, consider any two functions $F_1, F_2$, where $Q_i(y_2, w) = F_i(y_2, h(v)) = \alpha_i  y_2 + h(v), i = 1,2, \alpha_1 \neq \alpha_2, \alpha_1, \alpha_2 > 0$, and where $h(\cdot)$ is differentiable and satisfies homothetic separability. Then these functions satisfy Proposition \ref{prop:properties_F}: (i) take $y_2' > y_2, w$, then we have $Q_i(y_2', w) - Q_i(y_2, w) = \alpha_i (y_2' - y_2) > 0$; (ii) $h(v)$ is normalized to be homogeneous of degree one and hence concave and $\alpha_i \cdot y_2$ is concave, hence $F_i$ is concave and thus quasi-concave; (iii) is satisfied by the fact that $F_i$ is the sum of a linear function and a homogeneous function; (iv) is satisfied as for any $v \in V$, we can always find a $y_2$ such that $Q_i = y$ and the set is closed, and a $y_2'$ large enough such that the set is non-empty. They also satisfy Assumption \ref{asspt:properties_prodf}: (i) by homotheticity of $h(\cdot)$; (ii) by assumption; and (iii) can easily be shown by deriving the cost function from cost minimization. But, for a given $h(v)$, the set of functions $F_i$ of this form is infinitely large. Contradiction.
\end{proof}
\end{lemma}
\section{Additional Proofs} \label{sec_app:proofs}


\subsection{Proof of Corollary \ref{prop:revenue_ces_basic}}

Slightly rewriting Eq. \eqref{eq:CS_revenue} with e.g. $V = M$, I get,
\begin{equation} \label{eq-app:CS_revenue}
    r_{it} =   \sigma m_{it} + \frac{1-\sigma}{\sigma} \log \left( \frac{\beta_L}{\beta_M}  \left(\frac{L_{it}}{M_{it}}\right)^\sigma   + M_{it}^\sigma \right) +   \frac{\sigma-1}{\sigma} \log B - s_{it}^{*M} - \log \mathcal{E}_{it} + \varepsilon_{it},
\end{equation}
where $\bar{B} = (P^L_{it})^{\frac{\sigma}{\sigma-1}} \left(\frac{\beta_L}{\beta_M}\right)^{\frac{1}{\sigma-1}} + (P^M_{it})^{\frac{\sigma}{\sigma-1}}$.
 It is immediate that one can \textit{at most} identify $\sigma$ and the ratio $\frac{\beta_L}{\beta_M}$ from these revenue production functions. Non-identification of $v$ and $\omega_{it}$ is immediate. Non-identification of the output elasticities follows from the fact that these depend on $(1-\beta_L-\beta_M)$ as well (see Appendix \ref{sec_app:derivations}), which cannot be identified. $\blacksquare$

\section{Formal Derivation of Assumption \ref{asspt:scalar}} \label{sec_app:scalar}

Here, I formally derive the ``scalar unobservable'' assumption that demand for the freely variable inputs depends on a single (scalar) unobservable, $\omega_{it}$. I rely on notation and results introduced throughout the paper and do not reintroduce it for brevity's sake, but refer to the relevant results where required. 

The assumption states that
\[
V_{it} = S_t(\mathbf{X}_{it}, V_{-it}, \omega_{it}).
\]
From \eqref{eq:demirer} derived in the proof of Theorem \ref{thm:nonIdentification_K}, we have that,
\[
C\left( \tilde{Q}^*_{it}, \mathbf{X}_{it}, \textbf{P}^V_{it}, \omega_{it} \right) \coloneqq  F^{-1}\left( \mathbf{X}_{it}, \frac{\tilde{Q}^*_{it}}{\exp(\omega_{it})} \right)  C_2\left( \mathbf{X}_{it},  \mathbf{P}^V_{it} \right).
\]
By Shephard's lemma \citep{shephard1953cost}, the optimal input demand functions equal the derivative of the cost function with respect to the relevant input prices, that is,
\[
V^*_{it} = \dpar{C\left( \tilde{Q}^*_{it}, \mathbf{X}_{it}, \textbf{P}^V_{it}, \omega_{it} \right)}{P^V_{it}} =  F^{-1}\left( \mathbf{X}_{it}, \frac{\tilde{Q}^*_{it}}{\exp(\omega_{it})} \right)  \dpar{C_2\left( \mathbf{X}_{it},  \mathbf{P}^V_{it}\right)}{P^V_{it}}.
\]
Taking the ratio of any two freely variable inputs $V_{it}, V_{-it} \in \mathbf{V}_{it}$, we obtain,
\[
\frac{V^*_{it}}{V^*_{-it}} = \frac{\dpar{C_2\left( \mathbf{X}_{it},  \mathbf{P}^V_{it}\right)}{P^V_{it}}}{\dpar{C_2\left( \mathbf{X}_{it},  \mathbf{P}^V_{it}\right)}{P^{V-}_{it}}} \coloneqq \frac{C_V\left( \mathbf{X}_{it},  \mathbf{P}^V_{it}\right)}{C_{V-}\left( \mathbf{X}_{it},  \mathbf{P}^V_{it}\right)} \coloneqq C_{\mathbf{V}}\left( \mathbf{X}_{it},  \mathbf{P}^V_{it} \right).
\]
Then, assuming that input prices are common across firms, that is, $ \mathbf{P}^V_{it} =  \mathbf{P}^V_{t}$, we obtain,
\[
V^*_{it} =  C_{\mathbf{V}}\left( \mathbf{X}_{it},  \mathbf{P}^V_{t} \right) V_{-it} \coloneqq  S_t(\mathbf{X}_{it}, V_{-it}, \omega_{it}),
\]
where $\mathbf{P}^V_{t}$ is subsumed in the $t$ subscript of the $S_t$ function.

\section{Properties of the Production Function} \label{sec_app:production_function}

For ease of reference, I restate the standard assumptions that the production function $F(\cdot)$ and the production possibilities set need to satisfy. Denote $x$ the $n$-dimensional vector of inputs and $y$ the scalar output. When comparing two vectors $x,y$, let  $x\geq y$ indicate that all elements of $x$ are at least as great as the corresponding elements of $y$, and at least one element of $x$ is strictly greater than the corresponding elements of $y$. The production possibilities set $T$ is,
\begin{equation}
    T := \left\{ (x,y) : F(x) \geq y, x \geq 0 \right\},
\end{equation}
and it is assumed to satisfy \citep[p.252]{chambers1988applied},

\begin{assumption}[Properties of T] \label{asspt:properties_T} \
 \begin{enumerate}[(i)]
     \item T is nonempty
     \item T is a closed set
     \item T is a convex set
     \item if $(x,y) \in T, x^1 \geq x$, then $(x^1,y) \in T$ (free disposability of x)
     \item if $(x,y) \in T, y^1 \leq y$ then $(x,y^1) \in T$ (free disposability of y)
     \item for every finite $x$, $T$ is bounded from above
     \item $(x, 0) \in T$, but if $y \geq 0, (0_n, y) \notin T$ (weak essentiality),
 \end{enumerate}
\end{assumption}
\noindent where $0_n$ is the n-dimensional zero vector.

Based on these assumptions, the production function that is the solution to,
\begin{equation}
    F(x)  = \max\left\{ y : (x,y) \in T \right\},
\end{equation}
exists and has the following properties \citep{shephard2015theory, chambers1988applied}, 

\begin{proposition}[Properties of F] \label{prop:properties_F} \
    \begin{enumerate}[(i)]
        \item If $x' \geq x$ then $F(x') \geq F(x)$ (monotonicity)
        \item $V(y) \coloneqq \left\{ x : F(x) \geq y \right\}$ is a convex set (quasi-concavity)
        \item $F(0_n) = 0$ (weak essentiality)
        \item The set $V(y)$ is closed and non-empty for all $y > 0$.
    \end{enumerate}
\end{proposition}

\noindent Moreover, Proposition \ref{prop:properties_F} (iv) implies that the cost function exists \citep{mcfadden1978cost}. If, in addition, I assume that that $F(x)$ is finite, non-negative, and real-valued for all non-negative and finite $x$, then the cost function possesses standard properties \citep[p.52]{chambers1988applied}.

\section{Derivation of Parametric Revenue Production Functions} \label{sec_app:derivations}

\subsection{Cobb-Douglas Revenue Production Function}

The Cobb-Douglas cost minimization problem with $K_{it}$ as dynamic input and $L_{it}, M_{it}$ as variable inputs is,
\begin{align} \label{eq_app:cb_cost_min}
& \min_{L_{it}, M_{it}} P^L_{it} L_{it} + P^M_{it}  M_{it} \\ 
& \text{s.t. } K_{it}^{\beta_K} L_{it}^{\beta_L} M_{it}^{\beta_M} \text{exp}(\omega_{it}) \geq \tilde{Q}^{*}_{it} \nonumber
\end{align}
Taking first-order conditions and recombining them gives,
\begin{align}
L_{it} & = \frac{P^M_{it}}{P^L_{it}} M_{it} \\
M_{it} & = \frac{P^L_{it}}{P^M_{it}} L_{it}
\end{align}
Plugging these into the equality for $\tilde{Q}^*_{it}$ in turn and rewriting for the intermediate input functions gives, 
\begin{align}
L^*_{it} & = \left( \frac{\tilde{Q}^*_{it} \left( \frac{P^M_{it}}{P^L_{it}} \right)^{\beta_M}}{K_{it}^{\beta_K} \exp \omega_{it}} \right)^{\frac{1}{\nu}} \\
M^*_{it} & = \left( \frac{\tilde{Q}^*_{it} \left( \frac{P^L_{it}}{P^M_{it}} \right)^{\beta_L}}{K_{it}^{\beta_K} \exp \omega_{it}} \right)^{\frac{1}{\nu}},  
\end{align}
where we define $\nu \coloneqq \beta_L + \beta_M$ the short-term returns to scale. Plugging these back into the cost minimization objective function gives the Cobb-Douglas cost function,
\begin{equation} \label{eq_app:CD_CF}
    C_{\text{CD}}\left(K_{it}, \mathbf{P}^V_{it}, \frac{\tilde{Q}^*_{it}}{\exp(\omega_{it})}\right)= \frac{1}{\nu}  \left(\frac{\tilde{Q}^*_{it}}{\exp(\omega_{it})}\right)^{\frac{1-\nu}{\nu}}   K_{it}^{-\frac{\beta_K}{\nu}}  B,
\end{equation}
where the unit cost function is $B =  (P^L_{it})^{\frac{\beta_L}{\nu}} \cdot (P^M_{it})^{\frac{\beta_M}{\nu}}  \cdot \left( \left(\frac{\beta_M}{\beta_L}\right)^{\frac{\beta_L}{\nu}} + \left(\frac{\beta_L}{\beta_M}\right)^{\frac{\beta_M}{\nu}} \right) $. Hence, the marginal cost is,
\begin{equation} \label{eq_app:CD_MC}
    \lambda_{it}^{\text{CD}} = \frac{1}{\nu} \left( K_{it}^{\beta_K} L_{it}^{\beta_L} M_{it}^{\beta_M}\right)^{\frac{1}{\nu}-1 }K_{it}^{-\frac{\beta_K}{\nu}} \left( \frac{P^L_{it}}{P^M_{it}} \right)^{\frac{\beta_L - \beta_M}{\nu}} \mathcal{E}_{it}^{-1}.
\end{equation}
The output elasticities with respect to the variable inputs are $\beta_L$ and $\beta_M$. Plugging all of this into the RHS of the production function, we get Eq. \eqref{eq:CD_revenue}.

\subsection{CES Revenue Production Function}
The cost minimization problem is, 
\begin{align} \label{eq_app:ces_cost_min}
& \min_{L_{it}, M_{it}} P^L_{it} L_{it} + P^M_{it}  M_{it} \\ 
& \text{s.t. } \left( (1-\beta_L - \beta_M) K_{it}^\sigma + \beta_L  L_{it}^\sigma + \beta_M M_{it}^\sigma \right)^{\frac{v}{\sigma}} \exp(\omega_{it})  \geq \tilde{Q}^{*}_{it} \nonumber
\end{align}
Combining the first-order conditions gives,
\begin{align}
&L_{it} = \left( \frac{P^{L}_{it} \beta_M M_{it}^{\sigma-1}}{ P^M_{it} \beta_L  } \right)^\frac{1}{\sigma-1} \\ 
&M_{it} = \left( \frac{P^M_{it} \beta_L L_{it}^{\sigma-1} }{ P^L_{it} \beta_M }  \right)^{\frac{1}{\sigma - 1}}  
\end{align}
Plugging each of these into the equality for $\tilde{Q}^*_{it}$ and rewriting for the optimal inputs gives,
\begin{align}
L^*_{it} & = \left( \left(\frac{\tilde{Q}^*_{it}}{\exp(\omega_{it})}\right)^{\frac{\sigma}{v}} - \beta_K K_{it}^\sigma \right)^{\frac{1}{\sigma}} \left( \frac{\beta_L}{P^L_{it}} \right)^{-\frac{1}{\sigma-1}} B^{-\frac{1}{\sigma}} \\
M^*_{it} & = \left( \left(\frac{\tilde{Q}^*_{it}}{\exp(\omega_{it})}\right)^{\frac{\sigma}{v}} - \beta_K K_{it}^\sigma \right)^{\frac{1}{\sigma}} \left( \frac{1-\beta_L - \beta_M}{ P^M_{it} }\right)^{\frac{-1}{\sigma-1}} B^{-\frac{1}{\sigma}} 
\end{align}
where $B := (\pltilde)^{\frac{\sigma}{\sigma-1}} \beta_L^{-\frac{1}{\sigma-1}} + (P^M_{it})^{\frac{\sigma}{\sigma-1}} \beta_M^{-\frac{1}{\sigma-1}} $. Plugging these into the cost minimization objective function, the CES cost function is,
\begin{equation} 
C_{\text{CES}}\left(K_{it}, \mathbf{P}^V_{it}, \frac{\tilde{Q}^*_{it}}{\exp(\omega_{it})}\right) = \left( \left(\frac{\tilde{Q}^*}{\exp(\omega_{it})}\right)^{\frac{\sigma}{v}} - \beta_K K_{it}^\sigma \right)^{\frac{1}{\sigma}} B^{\frac{\sigma-1}{\sigma}},
\end{equation} Hence, the marginal cost is,
\begin{align}
    \lambda_{it}^{\text{CES}} & = \frac{1}{\sigma} \left( \left(\frac{Q^*_{it}}{\exp(\omega_{it})}\right)^{\frac{\sigma}{v}} - \beta_K K_{it}^{\sigma} \right)^{\frac{1}{\sigma}-1} \exp(\omega_{it})^{-1} B^{\frac{-1}{\sigma}} \left( (P^L_{it})^{\frac{\sigma}{\sigma-1}} \beta_L^{\frac{-1}{\sigma-1}} + (P^M_{it})^{\frac{\sigma}{\sigma-1}} \beta_M^{\frac{-1}{\sigma-1}} \right) \mathcal{E}_{it}^{-1} \nonumber \\
    & = \frac{1}{v} \left( \beta_L L_{it}^\sigma + \beta_M M_{it}^\sigma \right)^{\frac{1}{\sigma}-1} F(K_{it}, L_{it}, M_{it})^{\frac{\sigma}{v}-1} B^{\frac{\sigma-1}{\sigma}} \exp(\omega_{it})^{-1}  \mathcal{E}_{it}^{-1} ,
\end{align}
The output elasticity with respect to variable input $V_{it} \in \mathbf{V}_{it}$ is,
\begin{equation} \label{eq_app:CES_elasticity}
\dpar{f(\cdot)}{v_{it}} = v F(K_{it}, L_{it}, M_{it})^{\frac{-\sigma}{v}} \beta_V V_{it}^\sigma.
\end{equation}
Plugging all of this into the RHS of the production function and rewriting gives Eq. \eqref{eq:CS_revenue}.

\end{document}